\documentclass[submission,copyright,creativecommons]{eptcs}
\providecommand{\event}{Post Proceedings of ADG’23} 

\usepackage{graphicx}
\usepackage{cite}

\ifpdf
  \usepackage{underscore}         
  \usepackage[T1]{fontenc}        
\else
  \usepackage{breakurl}           
\fi

\title{Open Source Prover in the Attic}
\author{Zoltán Kovács
\institute{The Private University College of Education of the Diocese of Linz, Austria}
\email{zoltan.kovacs@ph-linz.at}
\and
Alexander Vujic
\institute{The Private University College of Education of the Diocese of Linz, Austria}
\email{alexander.vujic@ph-linz.at}
}
\def\titlerunning{Open Source Prover in the Attic}
\def\authorrunning{Z.~Kovács \& A.~Vujic}
\begin{document}
\maketitle

\begin{abstract}
The well known JGEX program became open source a few years ago, but seemingly, further development of the program can only be done without the original authors. In our project, we are looking at whether it is possible to continue such a large project as a newcomer without the involvement of the original authors. Is there a way to internationalize, fix bugs, improve the code base, add new features? In other words, to save a relic found in the attic and polish it into a useful everyday tool.
\end{abstract}

\section{Introduction}
Wide access to computers in education opens a new horizon in applying automated reasoning tools in the classroom. Many software tools exist already, that are of different levels of maturity, and several issues can be found that prevent direct use of them in education. One of the most mature tools is Java Geometry Expert (JGEX, \cite{Ye-Inproceedings,Ye2,Ye3}), which has its roots in Wichita State University and was developed by the automated geometry reasoning school, led by Chou and Gao, and programmed (mostly) by Ye.

JGEX is a remarkable summary of several decades of pioneering work \cite{Chou4}. It consists of more than 108,000 lines of Java code in more than 230 files. It has been open-sourced, however, just a couple of years ago (in 2016), and it was published on GitHub.\footnote{\url{https://github.com/yezheng1981/Java-Geometry-Expert}} Since then, more than 80 forks were created and the project was starred by more than 400 users. These numbers clearly show the popularity of JGEX, even if the original authors discontinued their work: Ye’s latest (very minor) change was committed in July 2018. Therefore, in our work, we raise the question, if it is still possible to contribute to this project significantly.

Why are further contributions necessary? First of all, the software tool supports only a couple of languages. For educational use, a translation for native speakers seems unavoidable: young learners may have difficulties with languages. Second, a modern user interface may be more appealing for the new generation of users. Third, science is clearly at a more mature state-of-the-art now, so some updates in the used algorithms seem important after a while. Fourth, it turned out that some minor, annoying bugs still exist in the program – they should be fixed someday: it is very important for a tool that is expected to be error-free in all matters when it is about theorem proving.

In our paper we discuss the possibility of continuing an existing tool for automated proving without the possibility to contact the original authors. Our discussion focuses on translating JGEX into German and Serbian, and on extending its capabilities to provide a better user experience. Also, we discuss how difficult it is to build the program with the newest development tools available, and what changes have to be done to be somewhat up-to-date in this concern.

We think this research is important for a whole community, in particular, to save the achieved algorithms and implementations \textit{for the next generation}. In other words, our research is a survival guide which aims at supporting the community for automated geometry reasoning.

\section{Expectations}

In this section we give an overview of the expected updates and further improvements.

\subsection{Translations}
JGEX’s final mainstream version supports the Chinese, English, Italian and Persian languages. Our research team, after forking the mainstream version in the repository,\footnote{\url{https://github.com/kovzol/Java-Geometry-Expert}} decided to add further translations: German, Portuguese and Serbian.

It turned out that the translation, technically speaking, is a simple process of creating a text file, similar to the CSV (comma separated values) format, that consists of a text ID integer, the English keyword (because in most cases it is used for the lookup, instead of the ID), the translated text, and (optionally) a tooltip. With minor edits, a new text file can easily be added to the source files and it can be inserted in the menu system.

Unfortunately, it also turned out that the tooltip entries are completely missing for all languages (except for English, but those translations are already given in the Java source code). On the other hand, some translations are completely missing by design: such expressions are short phrases like “by HYP” (which means: “by assumption”, “by hypothesis”) and “because”. We had to add such phrases in the CSV databases, and modify the Java source code to search for the phrases also in the CSV files.

Also, two additional lists of phrases are hardcoded in the original source code. They belong to the Geometry Deduction Database (GDD, \cite{Ye3,Chou5}) and the rule database for the Full Angle method [3]. These are given only in English. It is an ongoing work to move these phrases in the CSV database, instead of using the Java files to store the English version, and, in our fork, the German and Serbian translations as well.

Why are all translations stored in one CSV database? There would be a
more efficient way of translating the user interface into different
languages. Nevertheless, a variety of languages have already been
translated in the CSV database, therefore it was more accessible for the
team to continue working with CSV. In our case, the most extensive work
was dedicated to the Serbian translation. Our translation team, led by
the second author, started to make a copy of the German translation
(because of the better knowledge of German than English). After finishing
translation of the phrases in the CSV file, it turned out that, in
addition, both rule databases need to be translated. Then, some
additional phrases were found in the Java source code that had to be
translated too. To make a quality work, the translated phrases had to be
double-checked by native speakers who were also experts in geometry.
Unfortunately, by doing the translation in three different steps, every
participant had to work three times: once for the CSV entries, once for
the rule databases, and again for the unexpected phrases shown up
randomly in the source code. This made the work quite inefficient and
unexpectedly long.

Overall, it would have been better to unify the translation system first, and only after then start a concentrated work on doing the real effort in translating the phrases. But this was, actually, not really possible: JGEX has no full documentation, and it cannot be therefore assumed that any users (either newcomers or experienced ones) have a complete overview of all phrases used in the program.

We also learned that in many cases, unfortunately, the English translation contains spelling errors. As a consequence, the translation keys have spelling problems too. This makes it difficult to modify all erroneous translation keys in all CSV files and the Java source code at the same time.

In fact, there are sophisticated translation systems like \textit{gettext}. We will consider changing the translation system in a next round of possible updates.

\subsection{Modern Interface}

First author invited a group of prospective mathematics teachers at the Private University College of Education of the Diocese of Linz (PHDL) to give some feedback on the usability of JGEX. While most feedback was positive, several students mentioned that JGEX had an unusual user interface, and an old-fashioned look. Many features were difficult to find. Among others, to obtain a proof that is easy to understand in the classroom, the user needed to find an appropriate example in the database of showcases. The general method of showing the proof is to fix the objects and to select the relevant one from them, and use the right mouse click to access the proof. In certain cases, such as “08_9point.gex” the students may access the proof by right clicking on “SHOW: CYCLIC D G E F” and selecting “Prove”.

To solve such difficulties, several ideas were suggested by the students, and the authors of this paper. One simple possibility is to create tutorials in both textual and video format. On the other hand, a simplified view of the most useful capabilities for classroom use could be helpful. Here we refer to another software package, OK Geometry \cite{Magajna6}, which comes in different editions (Easy, Basic, Plus modes). Also, a more extensive use of tooltips could tear down some barriers. Another option is to force the user into a workflow: first, she had to create or load a construction, second, point to the searched properties, and third, to obtain the proof.

Today’s users are more familiar with web applications and mobile phone apps than native applications. In fact, the original version of JGEX comes with no installer: the user needs to download the source code, and it is her own task to compile and run it with an appropriate Java Development Kit (JDK) and Java Runtime Environment (JRE). First author maintains a ZIP package for Windows and Mac systems (based on the \textit{packr} utility\footnote{\url{https://github.com/libgdx/packr}}) by providing all required files to run JGEX without any further preparation.\footnote{\url{https://github.com/kovzol/Java-Geometry-Expert/releases}} Also, a Linux Snap version of JGEX is also available.\footnote{\url{https://snapcraft.io/jgex}} As of May 2023 there are 175 installations world-wide registered from 48 territories. Anyway, these packages are just one step towards simplifying the access to JGEX – we think it is unavoidable to make JGEX available as an embeddable web application by compiling its codebase to JavaScript.

In fact, Ye’s version of JGEX does not even compile automatically when the newest Java version is used. Some minor updates are required on the source code: Either the Java version must be downgraded to 8, or some small changes need to be done to avoid compiler errors. These requirements have been, luckily, also identified by other contributors who forked JGEX. On the other hand, the Java technology, selected by the Chinese experts, has proven to be a good choice, because with just minor modifications, it is easy to import the project into today’s favorite Java Integrated Desktop Environment (IDE), \textit{IntelliJ IDEA}.

Finally, we think it is unavoidable to make it possible to import \textit{GeoGebra} \cite{Hohernwart7} figures in JGEX, and to export them in GeoGebra format, since GeoGebra became the de facto standard of dynamic geometry during the last decade \cite{Thaller8}. Even if JGEX supports a wide set of drawing tools, for newcomers, it can be difficult to learn its toolset quickly enough.

\subsection{State-of-the-art Mathematics}

JGEX implements several mathematical algorithms. One of the supported algorithms is the Gröbner basis method \cite{Kapur9} that is known to be slower than the other methods, in a substantial set of input cases. Meanwhile, however, major speedups have been reported in some implementations of computing Gröbner bases. One of the successful implementations is included in GeoGebra’s embedded computer algebra system Giac \cite{Kovacs10}.

In fact, the Gröbner basis method, when using elimination, is known to provide better non-degeneracy conditions than the faster algebraic method, Wu’s approach. This result could also be incorporated into JGEX. It's important to note that in many cases, the degeneracy conditions output by Wu's method are necessary for the theorems to hold, and they help distinguish the cases when the theorem is true and when it is false. This result could also be incorporated into JGEX.

\subsection{Fixing Bugs and Adding Improvements}

Even for mature software tools, there is always room to improve minor problems. Among other minor issues, the proof protocol shown for the GDD method, could be improved by reducing the number of output lines and showing the hierarchy of the proof in addition. We show this concept below.

\begin{figure}[ht]
    \centering
    \includegraphics[width=\textwidth]{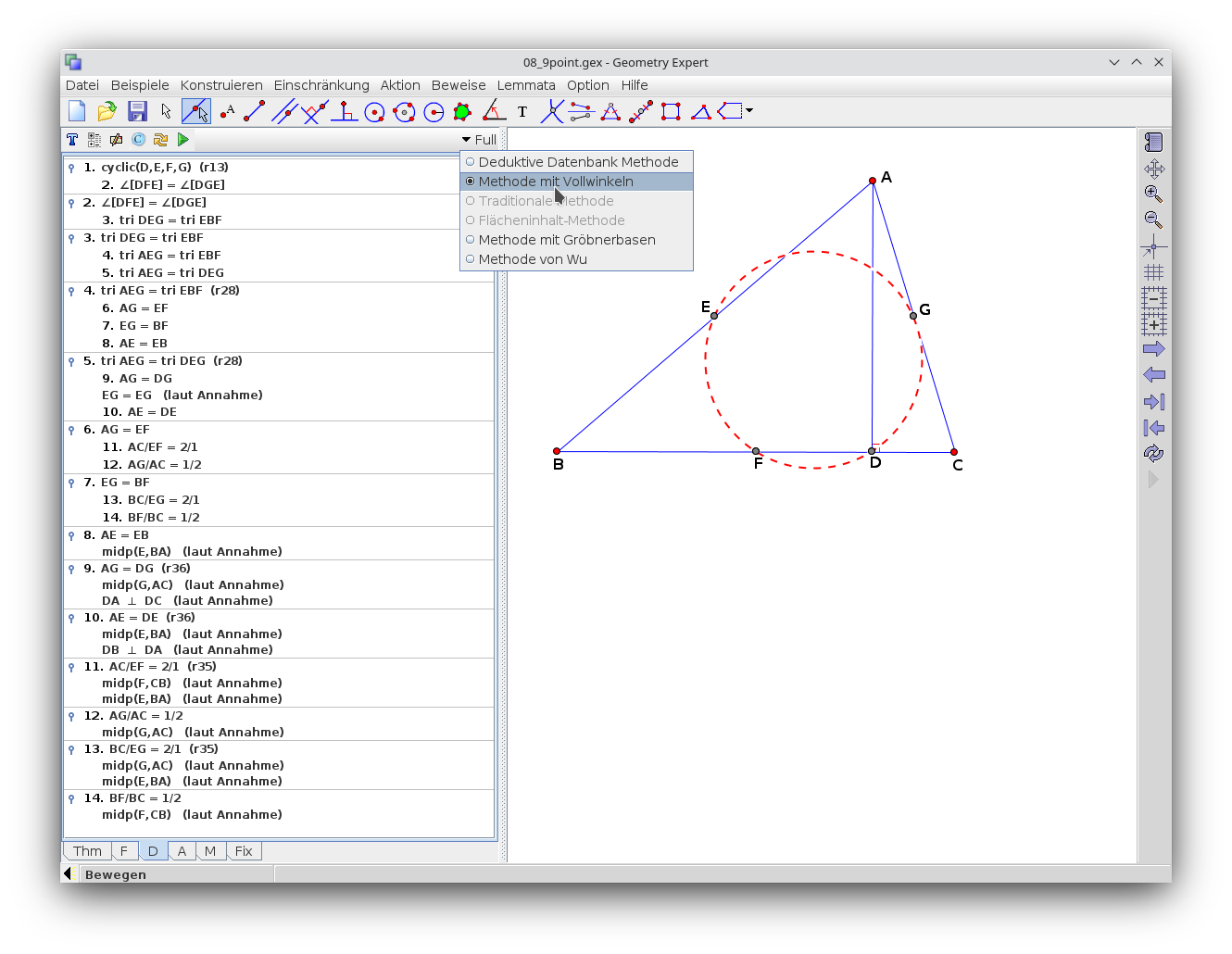}
    \caption{JGEX proves a part of the nine-point theorem by using the GDD method. Its user interface is set to German.}
     \label{fig:9point}
\end{figure}

We take the example \textbf{1_TOP_TEN/08_9point}: It constructs a figure to illustrate the nine-point circle theorem (\autoref{fig:9point}). We assume a classroom situation to get a readable proof of the fact that the three midpoints \textit{(E, F, G)} of an arbitrary triangle \textit{ABC} and a perpendicular foot point \textit{D} (of vertex \textit{A}, projected on side \textit{BC}) are concyclic.

\begin{figure}[ht]
    \centering
    \includegraphics[width= 200 px]{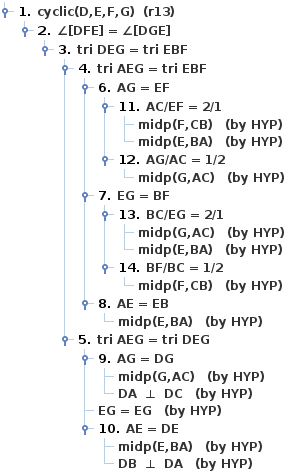}
    \caption{The output of the GDD method is visualized as a tree.}
    \label{fig:JGEX - Output of the GDD method}
\end{figure}

Now, by issuing some improvements on the source code (further details\footnote{\url{https://github.com/kovzol/Java-Geometry-Expert/commit/bc91b9ec916f97e38a100c08bec5bfda0c49de8d}}), we can communicate the proof in a simpler way (\autoref{fig:JGEX - Output of the GDD method}).

In fact, the change being performed to get this improvement is quite simple. Of course, one needs to have a deeper knowledge in the Java language and eventually in the mathematical background to achieve such changes in a feasible time. But, we can report that it is possible, and it does not require a high amount of time. Thus, a follow-up and continuation of the stopped work seems more than possible

\begin{figure}[ht]
    \centering
    \includegraphics[width=400 px]{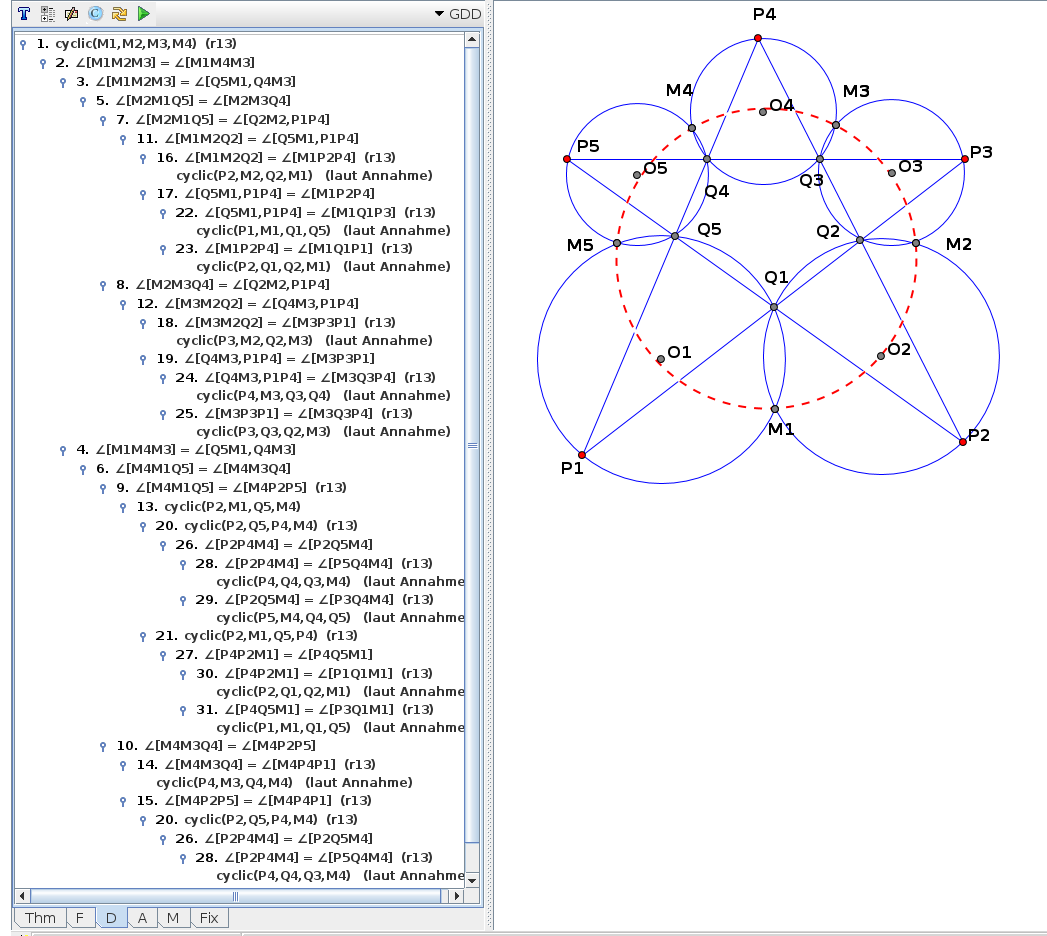}
    \caption{The GDD method finds a proof that is not a tree.
The interface is set to German and the connecting edges of the graph are not shown.}
     \label{fig:nonTree}
\end{figure}

\begin{figure}[ht]
    \centering
    \includegraphics[width=300 px]{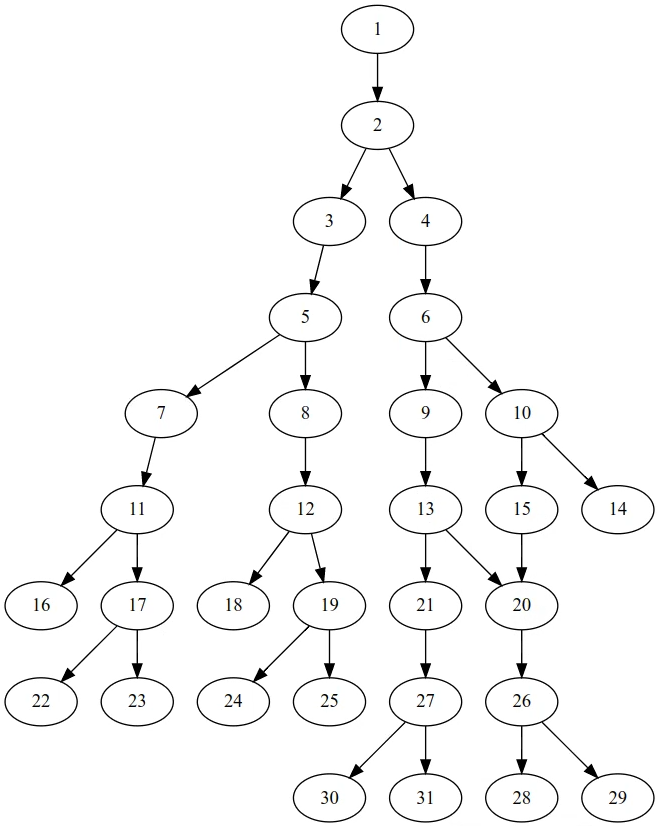}
    \caption{A better visualization of the GDD proof by using GraphViz.}
     \label{fig:Tree map - V1}
\end{figure}

Here we highlight that a tree structure for the obtained GDD proofs cannot always be accomplished. For example, \autoref{fig:nonTree} shows the example \textbf{1_TOP_TEN/10_5cir} which states the concyclicity of points \textit{M1}, \textit{M2}, \textit{M3} and \textit{M4} in the given figure. Here we can learn that node 20 is used in nodes 13 and 15 too, so the whole subtree of node 20 is repeated in the subtrees of nodes 13 and 15. The presence of this duplication may be inconvenient for some users. Therefore, an option to switch forcing the structure off, could be a solution, or, even better, a different way of visualization might be applied. One possibility to do that is the embedding of the GraphViz library,\footnote{\url{https://graphviz.org/}} to provide a professional look of the outline of the proof structure (\autoref{fig:Tree map - V1}).

Structured proofs may be beneficial when explaining the proof steps directly in the classroom. Of course, proofs that consist of a large number of steps, may be inappropriate for most students, but rather for gifted learners, mostly as preparations or training exercises for mathematical contests. Even so, providing a structured view of the proofs obtained by the GDD algorithm seems to be a substantial improvement for many classroom situations.

\section{Conclusion}

JGEX is an eminent software tool that summarizes several decades of pioneer work. Unfortunately, its main authors no longer maintain the code base. In our research we studied the question if it is possible to continue their work in some sense, and forward their legacy to the new generations.

According to our case study, we think that such a follow-up is possible, however, it is not straightforward. Several problems, mostly technical ones, can occur. The free availability of the source code is, however, a great help. With enough time and patience, the original version can be extended in a direction that seems fruitful for the long term.

We need to mention the quality of the code base. Unfortunately, the number of comments is very low. On the other hand, the variable names and the naming system for the Java methods are quite straightforward. That is, with little work, it is possible to learn the internals of the Java source code. Some parts of the code, however, require a major restructuring. Among others, the way of how the translations are handled, needs to be improved significantly. Also, some Java coding standards like capitalization of the source files, seem to be ignored. Nevertheless, by using modern refactoring tools, these issues could be solved with minimal efforts.

We remark that a general inspection process in IntelliJ IDEA 2023.1.2 found 269 errors, 9645 warnings, 430 weak warnings, 181 grammar errors and 16449 typos in JGEX’s code base. Even if many of the items of such reports can be of a matter of taste and coding style, a detailed study of these messages seems unavoidable for the long term.

Some recent work \cite{Beata11,Quaresma12,Teles13} confirm that continuing the pioneer work being put in JGEX is an important step towards the more general use of automated reasoning in the classroom. One possibility is to copy the algorithms and modify them accordingly (this way was chosen by Baeta and Quaresma, using C++), but another option (we chose this) is to use the original source code and do the modifications on it directly.

As a final conclusion, we think it is possible to save the heritage of the Chinese experts, and continue the hard work of popularizing JGEX and extending the user community of automated deduction in geometry.

\section{Acknowledgments}

The first author was partially supported by a grant PID2020-113192GB-I00 (Mathematical Visualization: Foundations, Algorithms and Applications) from the Spanish MICINN.

We acknowledge the kind support of translators Benedek Kovács, Alexander Thaller, Engelbert Zeintl (German), Jorge Cassio (Portuguese), Amela Hota, Predrag Janičić and Jelena Marković (Serbian).

We sincerely appreciate Amela Hota for her support in crafting a presentation of a preliminary version of this paper.

\bibliography{library}

\end{document}